# Charginos and Neutralinos Production at 3-3-1 Supersymmetric Model in $e^-e^-$ Scattering

M. Capdequi-Peyranère and M. C. Rodriguez[*]

*Physique Mathématique et Théorique, CNRS-UMR 5825*
*Université Montpellier II, F-34095, Montpellier Cedex 5.*

The goal of this article is to derive the Feynman rules involving charginos, neutralinos, double charged gauge bosons and sleptons in a 3-3-1 supersymmetric model. Using these Feynman rules we will calculate the production of a double charged chargino with a neutralino and also the production of a pair of single charged charginos , both in an electron- electron $e^-e^-$ process.

PACS numbers: 12.60.-i 12.60.Jv 13.85.Lg

## I. INTRODUCTION

The Standard Model is exceedingly successful in describing leptons, quarks and their interactions. Nevertheless, the Standard Model is not considered as the ultimate theory since neither the fundamental parameters, masses and couplings, nor the symmetry pattern are predicted. Even though many aspects of the Standard Model are experimentally supported to a very accuracy, the embedding of the model into a more general framework is to be expected. The possibility of a gauge symmetry based on the following symmetry $SU(3)_C \otimes SU(3)_L \otimes U(1)_N$ (3-3-1) [1] is particularly interesting, because it explains some fundamental questions that are eluded in the Standard Model [3].

Recently was proposed the supersymmetric extension of the 3-3-1 model [3]; in this kind of model the supersymmetry is naturally broken at a TeV scale, and the lightest scalar boson has an upper limit, which is 124.5 GeV for a given set of parameters at the tree level. We must remember that in MSSM the bound using radiative corrections to the mass of the lightest Higgs scalar is 130 GeV [4]. On another hand no direct observation of a Higgs boson has been made yet and current direct searches constraint its mass to $M_h > 113$ GeV [5] in the minimal Standard Model.

Linear colliders would be most versatile tools in experimental high energy physics. A large international effort is currently under way to study the technical feasibility and physics possibilities of linear $e^+e^-$ colliders in the TeV range. A number of designs have already been proposed (NLC, JLC, TESLA, CLIC, VLEPP, ...) and several workshops have recently been devoted to the subject. Not only they can provide $e^+e^-$ collisions and high luminosities, but also very energetic beams of real photons. One could thus exploit $\gamma\gamma$, $e^-\gamma$ and even $e^-e^-$ collisions for physics studies.

These exciting prospects have prompted a growing number of theoretical studies devoted to the investigation of the physics potential of such new $e^-e^-$ accelerator experiments. From the experimental point of view such a machine would provide relevant new possibilities, including the use of highly polarized beams or the production of high energy electron beam. Of course, in the realm of the Standard Model this option is not particularly interesting because mainly Møller scattering and bremsstrahlung events are to be observed. However, it is just for that reason that $e^-e^-$ collisions can provide crucial information on exotic processes, in particular on processes involving lepton and/or fermion number violation. Therefore, new perspectives emerge in detecting new physics beyond the Standard Model in processes having non-zero initial electric charge (and non-zero lepton number) like in electron-electron ($e^-e^-$) process.

In this paper we will explicitly work out two $e^-e^-$ processes and we will show on an example a difference between MSSM and the 3-3-1 supersymmetric model. This paper is organized as follows. The model is outlined in section II, where leptons, sleptons, scalars, higgsinos, gauge bosons and gauginos are defined. We derive the mass spectrum in section III, while the interactions are get in section IV. In order to give some examples, we derive the Feynman rules in the section V while the examples are worked out in section VI. Our conclusions are given in VII.

The lagrangian is given in Appendix A, while in Appendix B we show all non diagonal mass matrices of the charginos and neutralinos. The procedure to write two-components spinors in terms of four-components spinors is given in Appendix C. The differential cross sections of the processes we have calculated are given in the Appendix D.

---

[*]Permanent Address:Instituto de Física Teórica Universidade Estadual Paulista

## II. PARTICLE CONTENT

First of all, let us outline the model, following the notation given in [3]. We are writing here the particle content that we are using in this study, so it means we are not going to discuss quarks and squarks.

The leptons and sleptons are given by

$$L_l = \begin{pmatrix} \nu_l & l & l^c \end{pmatrix}_L^t, \quad \tilde{L}_l = \begin{pmatrix} \tilde{\nu}_l & \tilde{l} & \tilde{l}^c \end{pmatrix}_L^t, \quad l = e, \mu, \tau. \tag{2.1}$$

The scalars of our theory are

$$\eta = \begin{pmatrix} \eta^0 \\ \eta_1^- \\ \eta_2^+ \end{pmatrix}, \quad \rho = \begin{pmatrix} \rho^+ \\ \rho^0 \\ \rho^{++} \end{pmatrix}, \quad \chi = \begin{pmatrix} \chi^- \\ \chi^{--} \\ \chi^0 \end{pmatrix}, \quad S = \begin{pmatrix} \sigma_1^0 & \frac{h_2^+}{\sqrt{2}} & \frac{h_1^-}{\sqrt{2}} \\ \frac{h_2^+}{\sqrt{2}} & H_1^{++} & \frac{\sigma_2^0}{\sqrt{2}} \\ \frac{h_1^-}{\sqrt{2}} & \frac{\sigma_2^0}{\sqrt{2}} & H_2^{--} \end{pmatrix}, \tag{2.2}$$

the higgsinos, the superpartners of the scalars, are defined as

$$\tilde{\eta} = \begin{pmatrix} \tilde{\eta}^0 \\ \tilde{\eta}_1^- \\ \tilde{\eta}_2^+ \end{pmatrix}, \quad \tilde{\rho} = \begin{pmatrix} \tilde{\rho}^+ \\ \tilde{\rho}^0 \\ \tilde{\rho}^{++} \end{pmatrix}, \quad \tilde{\chi} = \begin{pmatrix} \tilde{\chi}^- \\ \tilde{\chi}^{--} \\ \tilde{\chi}^0 \end{pmatrix}, \quad \tilde{S} = \begin{pmatrix} \tilde{\sigma}_1^0 & \frac{\tilde{h}_2^+}{\sqrt{2}} & \frac{\tilde{h}_1^-}{\sqrt{2}} \\ \frac{\tilde{h}_2^+}{\sqrt{2}} & \tilde{H}_1^{++} & \frac{\tilde{\sigma}_2^0}{\sqrt{2}} \\ \frac{\tilde{h}_1^-}{\sqrt{2}} & \frac{\tilde{\sigma}_2^0}{\sqrt{2}} & \tilde{H}_2^{--} \end{pmatrix}. \tag{2.3}$$

To cancel chiral anomalies generated by these higgsinos we have to add the followings fields $\tilde{\eta}'$, $\tilde{\rho}'$, $\tilde{\chi}'$ and $\tilde{S}'$ and their respectives scalars, see [3].

When one breaks the 3-3-1 symmetry to the $SU(3)_C \otimes U(1)_{EM}$, the scalars get the following vacuum expectation values (VEVs):

$$<\eta> = \begin{pmatrix} v \\ 0 \\ 0 \end{pmatrix}, \quad <\rho> = \begin{pmatrix} 0 \\ u \\ 0 \end{pmatrix}, \quad <\chi> = \begin{pmatrix} 0 \\ 0 \\ w \end{pmatrix}, \quad <S> = \begin{pmatrix} 0 & 0 & 0 \\ 0 & 0 & \frac{z}{\sqrt{2}} \\ 0 & \frac{z}{\sqrt{2}} & 0 \end{pmatrix}, \tag{2.4}$$

and similar expressions to the prime fields, where $v = v_\eta/\sqrt{2}$, $u = v_\rho/\sqrt{2}$, $w = v_\chi/\sqrt{2}$, $z = v_{\sigma_2}/\sqrt{2}$, $v' = v_{\eta'}/\sqrt{2}$, $u' = v_{\rho'}/\sqrt{2}$, $w' = v_{\chi'}/\sqrt{2}$, and $z' = v_{\sigma_2'}/\sqrt{2}$.

In the 3-3-1 supersymmetric model the bosons of the symmetry $SU(3)_L$ are $V^a$ and their superpartners, the gauginos, are $\lambda_A^a$, with $a = 1, \cdots, 8$. For the $U(1)_N$ symmetry, we have the boson $V$ and its superpartner is written as $\lambda_B$ (that we call gaugino too).

In the usual 3-3-1 model [1] the gauge bosons are defined as

$$W_m^\pm(x) = -\frac{1}{\sqrt{2}}(V_m^1(x) \mp iV_m^2(x)), \quad V_m^\pm(x) = -\frac{1}{\sqrt{2}}(V_m^4(x) \pm iV_m^5(x)),$$

$$U_m^{\pm\pm}(x) = -\frac{1}{\sqrt{2}}(V_m^6(x) \pm iV_m^7(x)), \quad A_m(x) = \frac{1}{\sqrt{1+4t^2}}\left[(V_m^3(x) - \sqrt{3}V_m^8(x))t + V_m\right],$$

$$Z_m^0(x) = -\frac{1}{\sqrt{1+4t^2}}\left[\sqrt{1+3t^2}V_m^3(x) + \frac{\sqrt{3}t^2}{\sqrt{1+3t^2}}V_m^8(x) - \frac{t}{\sqrt{1+3t^2}}V_m(x)\right],$$

$$Z_m'^0(x) = \frac{1}{\sqrt{1+3t^2}}(V_m^8(x) + \sqrt{3}tV_m(x)), \tag{2.5}$$

where $t \equiv \tan\theta = \frac{g'}{g}$ and $g'$ and $g$ are the gauge coupling constants of $U(1)$ and $SU(3)$, respectively. We can define the charged gauginos, in analogy with the gauge bosons, in the following way

$$\lambda_W^\pm(x) = -\frac{1}{\sqrt{2}}(\lambda_A^1(x) \mp i\lambda_A^2(x)), \quad \lambda_V^\pm(x) = -\frac{1}{\sqrt{2}}(\lambda_A^4(x) \pm i\lambda_A^5(x)),$$

$$\lambda_U^{\pm\pm}(x) = -\frac{1}{\sqrt{2}}(\lambda_A^6(x) \pm i\lambda_A^7(x)). \tag{2.6}$$

## III. MASS SPECTRUM

The higgsino mass term comes from $\mathcal{L}_{HMT}$, see Eq.(A8), the mass of gauginos is given by $\mathcal{L}_{GMT}^{\text{gaugino}}$, see Eq.(A9) and the mixing term between gauginos and higgsinos is given by $\mathcal{L}_{H\tilde{H}\tilde{V}}^{\text{scalar}}$, Eq.(A6). After the mixture the physical states are double charged charginos, charginos and neutralinos. Now we will discuss these states.

### A. Double Charged Chargino

Doing the calculation we obtain the mass matrix of the double charged chargino in the following [1] way

$$\mathcal{L}_{\text{mass}}^{\text{double}} = -m_\lambda \lambda_U^{--} \lambda_U^{++} - \frac{\mu_\rho}{2} \tilde{\rho}'^{--} \tilde{\rho}^{++} - \frac{\mu_\chi}{2} \tilde{\chi}^{--} \tilde{\chi}'^{++} - \frac{\mu_S}{2} (\tilde{H}_1'^{--} \tilde{H}_1^{++} + \tilde{H}_2^{--} \tilde{H}_2'^{++})$$
$$- ig \left( u\tilde{\rho}^{++} - w'\tilde{\chi}'^{++} - \frac{z}{\sqrt{2}} \tilde{H}_1^{++} + \frac{z'}{\sqrt{2}} \tilde{H}_2'^{++} \right) \lambda_U^{--} - ig \left( w\tilde{\chi}^{--} - u'\tilde{\rho}'^{--} - \frac{z}{\sqrt{2}} \tilde{H}_2^{--} + \frac{z'}{\sqrt{2}} \tilde{H}_1'^{--} \right) \lambda_U^{++}$$
$$+ \frac{f_1 v}{3} \tilde{\chi}^{--} \tilde{\rho}^{++} - f_3 u \tilde{\chi}^{--} \tilde{H}_1^{++} - \sqrt{2} f_3 z \tilde{\chi}^{--} \tilde{\rho}^{++} - f_3 w \tilde{H}_2^{--} \tilde{\rho}^{++}$$
$$+ \frac{f_1' v'}{3} \tilde{\chi}'^{++} \tilde{\rho}'^{--} - f_3' u' \tilde{\chi}'^{++} \tilde{H}_1'^{--} - \sqrt{2} f_3' z' \tilde{\chi}'^{++} \tilde{\rho}'^{--} - f_3' w' \tilde{H}_2'^{++} \tilde{\rho}'^{--} + hc, \quad (3.1)$$

which can be writen in analogy with the MSSM[2], see Appendix B 1, as follows

$$\mathcal{L}_{\text{mass}}^{\text{double}} = -\frac{1}{2} \left( \Psi^{\pm\pm} \right)^t Y^{\pm\pm} \Psi^{\pm\pm} + hc. \quad (3.2)$$

The double chargino mass matrix is diagonalized using two rotation matrices, $A$ and $B$, defined by

$$\tilde{\chi}_i^{++} = A_{ij} \Psi_j^{++}, \quad \tilde{\chi}_i^{--} = B_{ij} \Psi_j^{--}, \quad i,j = 1, \cdots, 5. \quad (3.3)$$

where $A$ and $B$ are unitarity matrices choosen such that

$$M_{DCM} = B^* T A^{-1}. \quad (3.4)$$

To determine $A$ and $B$, we note that

$$M_{DCM}^2 = A T^t \cdot T A^{-1} = B^* T \cdot T^t (B^*)^{-1}, \quad (3.5)$$

which means that $A$ diagonalizes $T^t \cdot T$ while $B$ diagonalizes $T \cdot T^t$.

We define the following Dirac spinors to represent the mass eigenstates:

$$\Psi(\tilde{\chi}_i^{++}) = \left( \tilde{\chi}_i^{++} \;\; \bar{\tilde{\chi}}_i^{--} \right)^t, \quad \Psi^c(\tilde{\chi}_i^{--}) = \left( \tilde{\chi}_i^{--} \;\; \bar{\tilde{\chi}}_i^{++} \right)^t, \quad (3.6)$$

where $\tilde{\chi}_i^{++}$ is the particle and $\tilde{\chi}_i^{--}$ is the anti-particle( we are using the same notation as in [6]).

### B. Charged Chargino

We can write the mass matrix of the charged chargino in the following way

$$\mathcal{L}_{\text{mass}}^{\text{unique}} = -m_\lambda \left( \lambda_V^- \lambda_V^+ + \lambda_W^- \lambda_W^+ \right) - \frac{\mu_\eta}{2} \left( \tilde{\eta}_1^- \tilde{\eta}_1'^+ + \tilde{\eta}_2'^- \tilde{\eta}_2^+ \right) - \frac{\mu_\rho}{2} \tilde{\rho}'^- \tilde{\rho}^+ - \frac{\mu_\chi}{2} \tilde{\chi}^- \tilde{\chi}'^+$$
$$- \frac{\mu_S}{2} \left( \tilde{h}_1^- \tilde{h}_1'^+ + \tilde{h}_2'^- \tilde{h}_2^+ \right) + ig \left( v\tilde{\eta}_2^+ - w'\tilde{\chi}'^+ - \frac{z}{2} \tilde{h}_2^+ \right) \lambda_V^- + ig \left( w\tilde{\chi}^- - v'\tilde{\eta}_2'^- + \frac{z}{2} \tilde{h}_2'^- \right) \lambda_V^+$$
$$- ig \left( u\tilde{\rho}^+ - v'\tilde{\eta}_1'^+ + \frac{z}{2} \tilde{h}_1'^+ \right) \lambda_W^- - ig \left( v\tilde{\eta}_1^- - u'\tilde{\rho}'^- - \frac{z}{2} \tilde{h}_1^- \right) \lambda_W^+ - \frac{f_1 u}{3} \tilde{\chi}^- \tilde{\eta}_2^+ + \frac{f_1 w}{3} \tilde{\eta}_1^- \tilde{\rho}^+$$
$$- \frac{f_1' u'}{3} \tilde{\eta}_2'^- \tilde{\chi}'^+ + \frac{f_1' w'}{3} \tilde{\rho}'^- \tilde{\eta}_1'^+ - \frac{f_3 u}{\sqrt{2}} \tilde{\chi}^- \tilde{h}_2^+ - \frac{f_3 w}{\sqrt{2}} \tilde{h}_1^- \tilde{\rho}^+ - \frac{f_3' u'}{\sqrt{2}} \tilde{\chi}'^+ \tilde{h}_2'^- - \frac{f_3' w'}{\sqrt{2}} \tilde{h}_1'^+ \tilde{\rho}'^- + hc, \quad (3.7)$$

---

[1] Here we are considerating the conservation of $\mathcal{F}$ then $f_2, f_2' = 0$ [3].
[2] In the Appendix B we show all non diagonal mass matrices of the charginos and neutralinos

but Eq.(3.7), see Appendix B 2, takes the form

$$\mathcal{L}_{\text{mass}}^{\text{unique}} = -\frac{1}{2}\left(\Psi^{\pm}\right)^{t} Y^{\pm} \Psi^{\pm} + hc. \tag{3.8}$$

The chargino mass matrix is diagonalized using two rotation matrices, $D$ and $E$, defined by

$$\tilde{\chi}_i^+ = D_{ij}\Psi_j^+, \ \ \tilde{\chi}_i^- = E_{ij}\Psi_j^-, \ \ i,j=1,\cdots,8. \tag{3.9}$$

then we can write the diagonal mass matrix( $D$ and $E$ are unitary) as

$$M_{SCM} = E^* X D^{-1}. \tag{3.10}$$

As in the previous subsection we will define the following Dirac spinors:

$$\Psi(\tilde{\chi}_i^+) = \left(\begin{array}{cc} \tilde{\chi}_i^+ & \bar{\tilde{\chi}}_i^- \end{array}\right)^t, \ \ \Psi^c(\tilde{\chi}_i^-) = \left(\begin{array}{cc} \tilde{\chi}_i^- & \bar{\tilde{\chi}}_i^+ \end{array}\right)^t, \tag{3.11}$$

where $\tilde{\chi}_i^+$ is the particle and $\tilde{\chi}_i^-$ is the anti-particle. This structure is the same as is the chargino sector of the MSSM [6].

### C. Neutralino

For the neutralino case we have

$$\begin{aligned}\mathcal{L}_{\text{mass}}^{\text{neutralino}} = & -\frac{m_\lambda}{2}\left(\lambda_A^3 \lambda_A^3 + \lambda_A^8 \lambda_A^8\right) - \frac{m'}{2}\lambda_B \lambda_B - \frac{\mu_\eta}{2}\tilde{\eta}^0 \tilde{\eta}'^0 - \frac{\mu_\rho}{2}\tilde{\rho}^0 \tilde{\rho}'^0 - \frac{\mu_\chi}{2}\tilde{\chi}^0 \tilde{\chi}'^0 \\ & -\frac{\mu_S}{2}\left(\tilde{\sigma}_1'^0 \tilde{\sigma}_1^0 + \tilde{\sigma}_2'^0 \tilde{\sigma}_2^0\right) - \frac{ig'}{\sqrt{2}}(w\tilde{\chi}^0 - u\tilde{\rho}^0 + u'\tilde{\rho}'^0 - w'\tilde{\chi}'^0)\lambda_B \\ & -\frac{ig}{\sqrt{2}}\left(u\tilde{\rho}^0 - v\tilde{\eta}^0 + v'\tilde{\eta}'^0 - u'\tilde{\rho}'^0 + \frac{z}{2}\tilde{\sigma}_2^0 - \frac{z'}{2}\tilde{\sigma}_2'^0\right)\lambda_A^3 \\ & -\frac{ig}{\sqrt{6}}\left(2w\tilde{\chi}^0 - u\tilde{\rho}^0 - v\tilde{\eta}^0 + u'\tilde{\rho}'^0 + v'\tilde{\eta}'^0 - 2w'\tilde{\chi}'^0 + \frac{z}{2}\tilde{\sigma}_2^0 - \frac{z'}{2}\tilde{\sigma}_2'^0\right)\lambda_A^8 \\ & +\frac{f_1}{3}\left(u\tilde{\chi}^0 \tilde{\eta}^0 - v\tilde{\chi}^0 \tilde{\rho}^0 - w\tilde{\eta}^0 \tilde{\rho}^0\right) + \frac{f_1'}{3}\left(u'\tilde{\chi}'^0 \tilde{\eta}'^0 - v'\tilde{\chi}'^0 \tilde{\rho}'^0 - w'\tilde{\eta}'^0 \tilde{\rho}'^0\right) \\ & -\frac{f_3}{3}\left(u\tilde{\chi}^0 \tilde{\sigma}_2^0 + w\tilde{\rho}^0 \tilde{\sigma}_2^0 + 2z\tilde{\chi}^0 \tilde{\rho}^0\right) - \frac{f_3'}{3}\left(u'\tilde{\chi}'^0 \tilde{\sigma}_2'^0 + w'\tilde{\rho}'^0 \tilde{\sigma}_2'^0 + 2z'\tilde{\chi}'^0 \tilde{\rho}'^0\right) + hc \end{aligned} \tag{3.12}$$

Eq.(3.12) can be put in the following form, see Appendix B 3

$$\mathcal{L}_{\text{mass}}^{\text{neutralino}} = -\frac{1}{2}\left(\Psi^0\right)^t Y^0 \Psi^0 + hc. \tag{3.13}$$

The neutralino mass matrix is diagonalized by a $8 \times 8$ rotation matrix $N$, a unitary matrix satisfying

$$M_{NMD} = N^* Y^0 N^{-1}, \tag{3.14}$$

and the mass eigenstates are

$$\tilde{\chi}_i^0 = N_{ij}\Psi_j^0, \ \ j=1,\cdots,8. \tag{3.15}$$

We can define the following Majorana spinor to represent the mass eigenstates

$$\Psi(\tilde{\chi}_i^0) = \left(\begin{array}{cc} \tilde{\chi}_i^0 & \bar{\tilde{\chi}}_i^0 \end{array}\right)^t. \tag{3.16}$$

In Supersymmetric Left-Right models, there are double charged higgsinos too [7], and in both models the diagonalization of the charginos and neutralinos sectors is numerically performed.

## D. Sleptons

We can write the slepton mass term as

$$\mathcal{L}_{\text{mass}}^{\text{slepton}} = \mathcal{L}_{\text{mass}}^{\text{slep}} + \mathcal{L}_F^{\text{slepton}} + \mathcal{L}_D^{\text{slepton}}$$
$$= -\tilde{m}_\nu^2 \tilde{\nu}_l^* \tilde{\nu}_l - \left(\tilde{m}_L^2 + \frac{4m_l^2}{9}\right)\tilde{l}^*\tilde{l} - \left(\tilde{m}_R^2 + \frac{4m_l^2}{9}\right)\tilde{l}^{c*}\tilde{l}^c + \tilde{m}_{LR}^2(\tilde{l}\tilde{l}^c - \tilde{l}^{c*}\tilde{l}^*), \quad (3.17)$$

where

$$\tilde{m}_\nu^2 = m_\nu^2 + \frac{z^2 + 2v^2 - u^2 - w^2}{6},$$
$$\tilde{m}_L^2 = m_l^2 + \lambda_2^2 v^2 + \frac{z^2 - 2(w^2 - v^2 - u^2)}{12},$$
$$\tilde{m}_R^2 = m_{l^c}^2 + \lambda_2^2 v^2 + \frac{z^2 - 2(u^2 - 2w^2 - v^2)}{12},$$
$$\tilde{m}_{LR}^2 = \frac{9m_l}{4}\left[\left(\frac{\mu_S}{\sqrt{2}} + \frac{f_3 uw}{9z}\right) - \frac{4}{9}\zeta_0 z\right]. \quad (3.18)$$

Performing the diagonalization we find the following states

$$\begin{pmatrix} \tilde{l}_1^- \\ \tilde{l}_2^- \end{pmatrix} = \begin{pmatrix} \cos\theta_f & \sin\theta_f \\ -\sin\theta_f & \cos\theta_f \end{pmatrix} \begin{pmatrix} \tilde{l} \\ \tilde{l}^{c*} \end{pmatrix}, \quad (\tilde{l}_1^+ \; \tilde{l}_2^+) = (\tilde{l}^* \; \tilde{l}^c)\begin{pmatrix} \cos\theta_f & -\sin\theta_f \\ \sin\theta_f & \cos\theta_f \end{pmatrix}, \quad (3.19)$$

where the mixing angle is given by:

$$\tan 2\theta_f = \frac{2\tilde{m}_{LR}^2}{\tilde{m}_L^2 - \tilde{m}_R^2}, \quad (3.20)$$

and the masses of these states are:

$$m_{1,2}^2 = \frac{4m_l^2}{9} + \frac{1}{2}\left[\tilde{m}_L^2 + \tilde{m}_R^2 \pm \sqrt{(\tilde{m}_L^2 - \tilde{m}_R^2)^2 + 4\tilde{m}_{LR}^4}\right]. \quad (3.21)$$

Substituing the diagonal sleptons states, given in Eq.(3.19), in the Eq.(3.17), we obtain the following diagonal lagrangian

$$\mathcal{L}_{\text{mass}}^{\text{slepton}} = -\tilde{m}_\nu^2 \tilde{\nu}_l^* \tilde{\nu}_l - m_1^2 \tilde{l}_1^+ \tilde{l}_1^- - m_2^2 \tilde{l}_2^+ \tilde{l}_2^-. \quad (3.22)$$

We wish to stress that the slepton sector of this model is the same as in the MSSM [6].

## E. Neutral Gauge Boson

The neutral gauge boson mass is given by

$$\mathcal{L}_{\text{neutral}}^{\text{mass}} = (V_{3m} \; V_{8m} \; V_m)M^2\begin{pmatrix} V_3^m \\ V_8^m \\ V^m \end{pmatrix}, \quad (3.23)$$

where

$$M^2 = \frac{g^2}{2}\begin{pmatrix} (v^2 + u^2 + z^2 + 4z_1^2) & \frac{1}{\sqrt{3}}(v^2 - u^2 + z^2 + 4z_1^2) & -2tu^2 \\ \frac{1}{\sqrt{3}}(v^2 - u^2 + z^2 + 4z_1^2) & \frac{1}{3}(v^2 + u^2 + 4w^2 + z^2 + 4z_1^2) & \frac{2t}{\sqrt{3}}(u^2 + 2w^2) \\ -2tu^2 & \frac{2t}{\sqrt{3}}(u^2 + 2w^2) & 4t^2(u^2 + w^2) \end{pmatrix}, \quad (3.24)$$

where $t = g/g'$.

In the approximation that $w^2 \gg v^2, u^2, z^2$, the masses of the neutral gauge bosons are: 0, $M_Z^2$ and $M_{Z'}^2$, and the masses are given by

$$M_Z^2 \approx \frac{1}{2}\frac{g^2 + 4g'^2}{g^2 + 3g'^2}(v^2 + u^2 + z^2 + v'^2 + u'^2 + z'^2), \quad M_{Z'}^2 \approx \frac{1}{3}(g^2 + 3g'^2)(2w^2 + 2w'^2). \tag{3.25}$$

Using $M_W$ given in Eq.(3.29) and $M_Z$ in Eq.(3.25) we get the following relation:

$$\frac{M_Z^2}{M_W^2} = \frac{1 + 4t^2}{1 + 3t^2} = \frac{1}{1 - \sin^2\theta_W}, \tag{3.26}$$

therefore we obtain

$$t^2 = \frac{\sin^2\theta_W}{1 - 4\sin^2\theta_W}. \tag{3.27}$$

### F. Charged Gauge Boson

The gauge mass term is given by $\mathcal{L}_{HHVV}^{\text{scalar}}$, see Eq.(A6), which we can divided in $\mathcal{L}_{\text{charged}}^{\text{mass}}$ and $\mathcal{L}_{\text{neutral}}^{\text{mass}}$.
The charged gauge boson mass is

$$\mathcal{L}_{\text{charged}}^{\text{mass}} = M_W^2 W_m^- W^{+m} + M_V^2 V_m^- V^{+m} + M_U^2 U_m^{--} U^{++m}, \tag{3.28}$$

where

$$M_U^2 = \frac{g^2}{4}(v_\rho^2 + v_\chi^2 + 4v_{\sigma_2}^2 + v_{\rho'}^2 + v_{\chi'}^2 + 4v_{\sigma'_2}^2),$$

$$M_W^2 = \frac{g^2}{4}(v_\eta^2 + v_\rho^2 + 2v_{\sigma_2}^2 + v_{\eta'}^2 + v_{\rho'}^2 + 2v_{\sigma'_2}^2),$$

$$M_V^2 = \frac{g^2}{4}(v_\eta^2 + v_\chi^2 + 2v_{\sigma_2}^2 + v_{\eta'}^2 + v_{\chi'}^2 + 2v_{\sigma'_2}^2). \tag{3.29}$$

We want to mention that the gauge boson sector is exactly the same as in the non-supersymmetric 3-3-1 model [1].

### IV. INTERACTIONS

In the previous section we have analysed the physical spectrum of the model. Now we are going to get interactions between these particles.

The procedure to write two component spinors in terms of four components spinors is given in Appendix C.

### A. Lepton Gauge Boson Interaction

We will define the following Dirac spinors for the leptons

$$\Psi(l) = \begin{pmatrix} l & \bar{l}^c \end{pmatrix}^t, \quad \Psi^c(l) = \begin{pmatrix} l^c & \bar{l} \end{pmatrix}^t, \quad \Psi(\nu_l) = \begin{pmatrix} \nu_l & 0 \end{pmatrix}^t. \tag{4.1}$$

Due to Eqs.(4.1), we can rewrite Eq.(A3) as follows

$$\mathcal{L}_{llV}^{lep} = -\frac{g}{\sqrt{2}}\left(\bar{\Psi}^c(l)\gamma^m L\Psi(l)U_m^{++} + \bar{\Psi}^c(l)\gamma^m L\Psi(\nu_l)V_m^+ + \bar{\Psi}(\nu_l)\gamma^m L\Psi(l)W_m^+ + hc\right)$$

$$- \frac{gt}{\sqrt{1+4t^2}}\bar{\Psi}(l)\gamma^m\Psi(l)A_m - \frac{g}{2}\sqrt{\frac{1+4t^2}{1+3t^2}}\bar{\Psi}(\nu_l)\gamma^m L\Psi(\nu_l)\left[Z_m - \frac{1}{\sqrt{3}}\frac{1}{\sqrt{1+4t^2}}Z'_m\right]$$

$$- \frac{g}{4}\sqrt{\frac{1+4t^2}{1+3t^2}}\left[\left(\frac{-1}{1+4t^2}\bar{\Psi}(l)\gamma^m\Psi(l) + \bar{\Psi}(l)\gamma^m\gamma^5\Psi(l)\right)Z_m\right.$$

$$\left.+ \left(\frac{-\sqrt{3}}{\sqrt{1+4t^2}}\bar{\Psi}(l)\gamma^m\Psi(l) - \frac{1}{\sqrt{3}}\frac{1}{\sqrt{1+4t^2}}\bar{\Psi}(l)\gamma^m\gamma^5\Psi(l)\right)Z'_m\right] = \mathcal{L}_{llV}^{NC} + \mathcal{L}_{llV}^{CC}, \tag{4.2}$$

where $t$ is defined in Eq.(3.27).

Let us define also the following parameters

$$e = \frac{g \sin\theta}{\sqrt{1 + 3\sin^2\theta}}, \ h(t) = 1 + 4t^2, \ v_l = -\frac{1}{h(t)}, \ a_l = 1, \ v'_l = -\frac{\sqrt{3}}{\sqrt{h(t)}}, \ a'_l = \frac{v'_l}{3}, \ g^2 = \frac{8 G_F M_W^2}{\sqrt{2}}. \quad (4.3)$$

Then we can now rewrite the neutral part of Eq.(4.2):

$$\mathcal{L}_{llV}^{NC} = -e\bar{\Psi}(l)\gamma^m \Psi(l) A_m - \frac{g}{2}\frac{M_Z}{M_W}\bar{\Psi}(\nu_l)\gamma^m L \Psi(\nu_l) \left[ Z_m - \frac{1}{\sqrt{3}}\frac{1}{\sqrt{h(t)}} Z'_m \right]$$
$$- \frac{g}{4}\frac{M_Z}{M_W} \left[ \bar{\Psi}(l)\gamma^m (v_l + a_l\gamma^5)\Psi(l) Z_m + \bar{\Psi}(l)\gamma^m (v'_l + a'_l\gamma^5)\Psi(l) Z'_m \right], \quad (4.4)$$

The charged part of Eq.(4.2) is

$$\mathcal{L}_{llV}^{CC} = -\frac{g}{\sqrt{2}} \left( \bar{\Psi}^c(l)\gamma^m L \Psi(l) U_m^{++} + \bar{\Psi}^c(l)\gamma^m L \Psi(\nu_l) V_m^+ + \bar{\Psi}(\nu_l)\gamma^m L \Psi(l) W_m^+ + hc \right), \quad (4.5)$$

where $g$ is defined in Eq.(4.3). The lagrangian in the Eqs.(4.4, 4.5) is the same that it was shown in Ref [1].

### B. Chargino(Neutralino) Bilepton $U^{--}$ Interaction

Using Eq.(C2) in the Eqs(A5) and (A6), we can write the interaction between the double charged vector with the double charged chargino and the neutralino, in the following way

$$\mathcal{L}_{U\tilde{\chi}^{++}\tilde{\chi}^0} = -\frac{g}{2} \left\{ \left[ \bar{\tilde{W}}_3 L\gamma^m R\tilde{U} - \bar{\tilde{U}}^c L\gamma^m R\tilde{W}_3 + \sqrt{3}\left( \bar{\tilde{U}}^c L\gamma^m R\tilde{W}_8 - \bar{\tilde{W}}_8 L\gamma^m R\tilde{U} \right) \right. \right.$$
$$+ \sqrt{2}\left( \bar{\tilde{T}}_5^0 L\gamma^m R\tilde{T}_2^{++} + \bar{\tilde{T}}_4^0 L\gamma^m R\tilde{T}_1^{++} + \bar{\tilde{T}}_2^{c++} L\gamma^m R\tilde{T}_6^0 + \bar{\tilde{T}}_1^{c++} L\gamma^m R\tilde{T}_3^0 \right)$$
$$\left. + \bar{\tilde{S}}_2^{c++} L\gamma^m R\tilde{S}_4^0 + \bar{\tilde{S}}_1^{c++} L\gamma^m R\tilde{S}_3^0 + \bar{\tilde{S}}_4^0 L\gamma^m R\tilde{S}_1^{++} + \bar{\tilde{S}}_3^0 L\gamma^m R\tilde{S}_2^{++} \right] U_m^{--} + hc \right\}$$
$$= -\frac{g}{2} \left\{ \left[ \left( (N_{i1}^* - \sqrt{3}N_{i2}^*)B_{j1} + \sqrt{2}(N_{i8}^*B_{j3} + N_{i7}^*B_{j2}) + N_{i12}^*B_{j5} + N_{i13}^*B_{j4} \right) \bar{\Psi}(\tilde{\chi}_i^0) L\gamma^m R\Psi(\tilde{\chi}_j^{++}) \right. \right.$$
$$\left. \left. + \left( A_{i1}^*(\sqrt{3}N_{j2} - N_{j1}) + \sqrt{2}(A_{i3}^*N_{j9} + A_{i2}^*N_{j6}) + A_{i5}^*N_{j13} + A_{i4}^*N_{j12} \right) \bar{\Psi}^c(\tilde{\chi}_i^{--}) L\gamma^m R\Psi(\tilde{\chi}_j^0) \right] U_m^{--} + hc \right\}. \quad (4.6)$$

In a similar way we get for the charginos

$$\mathcal{L}_{U\tilde{\chi}^+\tilde{\chi}^+} = -\frac{g}{\sqrt{2}} \left[ \left( \bar{\tilde{W}}^c L\gamma^m R\tilde{V} - \bar{\tilde{V}}^c L\gamma^m R\tilde{W} + \bar{\tilde{T}}_2^{c+} L\gamma^m R\tilde{T}_1^+ + \bar{\tilde{T}}_1^{c+} L\gamma^m R\tilde{T}_2^+ \right. \right.$$
$$\left. \left. + \frac{1}{2}\bar{\tilde{S}}_1^{c+} L\gamma^m R\tilde{S}_2^+ + \frac{1}{2}\bar{\tilde{S}}_2^{c+} L\gamma^m R\tilde{S}_1^+ \right) U_m^{--} + hc \right]$$
$$= \frac{g}{\sqrt{2}} \left[ \left( D_{i1}^*E_{j2} - D_{i2}^*E_{j1} + D_{i4}^*E_{j3} + D_{i3}^*E_{j4}\frac{1}{2}D_{i7}^*E_{j8} + \frac{1}{2}D_{i8}^*E_{j7} \right) \bar{\Psi}^c(\tilde{\chi}_i^-) L\gamma^m R\Psi(\tilde{\chi}_j^+) U_m^{--} + hc \right]. \quad (4.7)$$

### C. Lepton Slepton Chargino(Neutralino) Interaction

The interaction between lepton slepton chargino(neutralino) is given by

$$\mathcal{L}_{ll\tilde{\chi}} = \left\{ -g\cos\theta_f A_i \bar{\Psi}(\tilde{\chi}_i^0) L\Psi(l) - g\sin\theta_f A_{i1}\bar{\Psi}(l)R\Psi^c(\tilde{\chi}_i^{--}) + 2\lambda_3 \sin\theta_f \left[ A_{i5}^* \bar{\Psi}^c(\tilde{\chi}_i^{--})L\Psi^c(l) + N_{i8}^*\bar{\Psi}(\tilde{\chi}_i^0)L\Psi(l) \right] \right\} \tilde{l}_1^+$$
$$+ \left\{ g\sin\theta_f A_i^*\bar{\Psi}(\tilde{\chi}_i^0)L\Psi(l) - g\cos\theta_f A_{i1}\bar{\Psi}(l)R\Psi^c(\tilde{\chi}_i^{--}) + 2\lambda_3\cos\theta_f \left[ A_{i5}^*\bar{\Psi}^c(\tilde{\chi}_i^{--})L\Psi^c(l) + N_{i8}^*\bar{\Psi}(\tilde{\chi}_i^0)L\Psi(l) \right] \right\} \tilde{l}_2^+$$
$$+ \left\{ -g\frac{2\lambda_3}{\sqrt{2}}\left[ E_{i8}^*\bar{\Psi}(\tilde{\chi}_i^+)L\Psi(l) + D_{i7}^*\bar{\Psi}^c(\tilde{\chi}_i^-)L\Psi^c(l) + \sqrt{2}N_{i7}^*\bar{\Psi}(\tilde{\chi}_i^0)L\Psi(\nu_l) \right] \right\} \tilde{\nu}_l + hc, \quad (4.8)$$

where $A_i = \left( \frac{N_{i1}^*}{\sqrt{2}} + \frac{N_{i2}^*}{\sqrt{6}} \right)$.

## V. FEYNMAN RULES

In the previous section we got the interaction lagrangians, mainly the following vertices:

- Lepton-Lepton-Gauge Boson,
- Chargino-Chargino(Neutralino)-Gauge Boson,
- Slepton-Lepton-Chargino(Neutralino).

In the Table I we give the Feynman rules for the interactions mentioned above.

In Table I we have defined the following operators:

$$O^1_{ij} = A^*_{i1}(\sqrt{3}N_{j2} - N_{j1}) + \sqrt{2}(A^*_{i3}N_{j9} + A^*_{i2}N_{j6}) + A^*_{i5}N_{j13} + A^*_{i4}N_{j12}; \tag{5.1}$$

$$O^2_{ij} = -\left(D^*_{i1}E_{j2} - D^*_{i2}E_{j1} + D^*_{i4}E_{j3} + D^*_{i3}E_{j4}\frac{1}{2}D^*_{i7}E_{j8} + \frac{1}{2}D^*_{i8}E_{j7}\right). \tag{5.2}$$

Because there are neutralinos, charginos, sleptons and sneutrinos in the MSSM, we show in Table II two different solutions got in mSUGRA [9]. To derive the total and differential cross sections, we will assume that the mass of the lightest particle of this model coincides with the mass parameter coming from MSSM and shown in Table II.

## VI. APPLICATIONS

In this section, we will perform the calculation of differential and total cross sections of the lightest chargino, double charged chargino and neutralinos, in $e^-e^-$ scattering. First we present the chargino production.

In the two subsections below, we neglect the electron mass and we assume that electrons have energy $E/2$. We will consider that $P_1$ and $P_2$ are the four momenta of the incoming electrons while $K_1$ and $K_2$ are the four momenta of the outgoing particles.

### A. Charginos Production $e^-e^- \to \tilde{\chi}^-\tilde{\chi}^-$

Considering the Table I, we have 5 diagrams at the tree level corresponding to this process, see Fig.(1). The first and the third diagrams that appear in Fig.(1) exist in the MSSM, but the other ones are new contributions coming from the 3-3-1 supersymmetric model.

We have calculated the differential cross section, see Eq.(D1), and the total cross section, and we have displayed several plots of the total cross section with $M_U \geq 0.5$TeV and $\sqrt{s} = 0.5, 1.0$ and $2.0$TeV.

The plots show that outside the $U$ resonance, the total cross section is of order of pb, like in the MSSM [10]. This result is diplayed in Figs.(2,3,4).

### B. Double Chargino and Neutralino Production $e^-e^- \to \tilde{\chi}^{--}\tilde{\chi}^0$

Considering the Table I, we have at the tree level 9 diagrams for this process, see Fig.(5). In Supersymmetric Left-Right model [7] the production of double charged higgsino occurs via a selectron exchange in t-channel, like in the third diagram of Fig.(5).

As in the previous subsection, we have calculated the differential cross section, see Eq.(D2), and the total cross section. We have done several plots using $M_{\chi^0}$, $M_{\tilde{l}_1}$ and $M_{\tilde{l}_2}$, given in Table II, and $M_U = 0.5$TeV. Some of our results are show in Figs.(6,7,8).

## VII. CONCLUSIONS

Because of low level of Standard Model backgrounds, the total cross section $\sigma \approx 10^{-3}nb$ at $\sqrt{s} = 500GeV$ [12], $e^-e^-$ collisions are a good reaction for discovering and investigating new physics at linear colliders.

We have shown in this work that the production of single charginos have more contributions in this model than in the MSSM. Although in the MSSM the chargino pairs can be only produced through $e^-e^-$ collisions by sneutrino exchanges in u and t channels , in the 3-3-1 supersymmetric model we have also a s channel contribution due to the exchange of a bilepton $U^{--}$. This new contribution induces a peak at $\sqrt{s} \simeq M_U$, where $M_U$ is the bilepton mass, and gives an clean signal. Near the resonance the dominant term in the total cross section is given by $|O_2|^2(m^4_{\tilde{\chi}^+} - 8sm^2_{\tilde{\chi}^+} + 4s^2)$ which is a clear signal coming from the bilepton $U$ contribution. The total cross section outside the $U$'s resonance has the same order of magnitude than the cross section in the MSSM, as we should expected, because in this case we don't have an enhancement due to the $s$- channel contribution.

It was shown too that in this model we have double charged charginos, that is impossible in the MSSM framework. Therefore it is a very useful way to distinguish the 3-3-1 supersymmetric model from the MSSM, and also from the usual 3-3-1 model because in this case we don't have double charged leptons. We have considered the double chargino mass in the range from $700 \leq M_{\tilde{\chi}^{++}} \leq 800$ GeV, and we could get cross section of the order of pb outside the $U$ ressonance, while in the bilepton ressonance we have an enhance in the cross section. We believe that these new states can be discovered, if they really exist, in the linear colliders as (NLC, JLC, TESLA, CLIC, VLEPP, ...).

## ACKNOWLEDGMENTS

This research was financed by Fundação de Amparo à Pesquisa do Estado de São Paulo (M.C.R.). One of us (M.C.R.) would like to thank the Laboratoire de Physique Mathématique et Théorique (Montpellier) for it kind hospitality and also G. Moultaka for useful discussions.

## APPENDIX A: LAGRANGIAN

With the fields introduced in section II, we can built the following lagrangian [3]

$$\mathcal{L}_{331S} = \mathcal{L}_{SUSY} + \mathcal{L}_{\text{soft}} , \tag{A1}$$

where

$$\mathcal{L}_{SUSY} = \mathcal{L}_{\text{Lepton}} + \mathcal{L}_{\text{Quarks}} + \mathcal{L}_{\text{Gauge}} + \mathcal{L}_{\text{Scalar}}, \tag{A2}$$

is the supersymmetric part while $\mathcal{L}_{\text{soft}}$ breaks supersymmetry. Now we are going to present all the lagrangian of the model, except the quark part, because we are not considering them in this study.

### 1. Lepton Lagrangian

In the $\mathcal{L}_{Lepton}$ term we have the interaction between leptons and gauge bosons in component are given by

$$\mathcal{L}^{lep}_{llV} = \frac{g}{2}\bar{L}\bar{\sigma}^m \lambda^a L V^a_m, \tag{A3}$$

where $\lambda^a$ are the usual Gell-Mann Matrices. The next part, we are interested here, is the interaction between lepton-slepton-gaugino is given by the following term

$$\mathcal{L}^{lep}_{l\tilde{l}\tilde{V}} = -\frac{ig}{\sqrt{2}}(\bar{L}\lambda^a \tilde{L}\bar{\lambda}^a_A - \bar{\tilde{L}}\lambda^a L \lambda^a_A). \tag{A4}$$

### 2. Gauge Lagrangian

The part we are interested in $\mathcal{L}_{Gauge}$ is the interaction between higgsino and gauge boson, that is:

$$\mathcal{L}^{\text{gauge}}_{\lambda\lambda V} = -ig f^{abc} \bar{\lambda}^a_A \lambda^b_A \sigma^m V^c_m, \tag{A5}$$

$f^{abc}$ are the structure constants of the gauge group $SU(3)$.

### 3. Scalar Lagrangian

While in the scalar sector we have the following three terms, considering all triplets and anti-sextet:

$$\begin{aligned}
\mathcal{L}_{H\tilde{H}V}^{\text{scalar}} &= -\frac{ig}{\sqrt{2}}\left[\bar{\tilde{\eta}}\lambda^a\eta\bar{\lambda}_A^a - \bar{\eta}\lambda^a\tilde{\eta}\lambda_A^a + \bar{\tilde{\rho}}\lambda^a\rho\bar{\lambda}_A^a - \bar{\rho}\lambda^a\tilde{\rho}\lambda_A^a + \bar{\tilde{\chi}}\lambda^a\chi\bar{\lambda}_A^a - \bar{\chi}\lambda^a\tilde{\chi}\lambda_A^a + \bar{\tilde{S}}\lambda^a S\bar{\lambda}_A^a \right. \\
&\quad \left. - \bar{S}\lambda^a\tilde{S}\lambda_A^a\right]V_m^a - \frac{ig'}{\sqrt{2}}\left[\bar{\tilde{\rho}}\rho\bar{\lambda}_B - \bar{\rho}\tilde{\rho}\lambda_B - \bar{\tilde{\chi}}\chi\bar{\lambda}_B + \bar{\chi}\tilde{\chi}\lambda_B\right]V_m, \\
\mathcal{L}_{\tilde{H}\tilde{H}V}^{\text{scalar}} &= \frac{g}{2}\left[\bar{\tilde{\eta}}\bar{\sigma}^m\lambda^a\tilde{\eta} + \bar{\tilde{\rho}}\bar{\sigma}^m\lambda^a\tilde{\rho} + \bar{\tilde{\chi}}\bar{\sigma}^m\lambda^a\tilde{\chi} + \bar{\tilde{S}}\bar{\sigma}^m\lambda^a\tilde{S}\right]V_m^a + \frac{g'}{2}\left[\bar{\tilde{\rho}}\bar{\sigma}^m\tilde{\rho} - \bar{\tilde{\chi}}\bar{\sigma}^m\tilde{\chi}\right]V_m, \\
\mathcal{L}_{HHVV}^{\text{scalar}} &= \frac{1}{4}\left[g^2 V_m^a V^{bm}\bar{\eta}\lambda^a\lambda^b\eta + g^2 V_m^a V^{bm}\bar{\rho}\lambda^a\lambda^b\rho + g^2 V_m^a V^{bm}\bar{\chi}\lambda^a\lambda^b\chi \right. \\
&\quad + g^2 V_m^a V^{bm}\left(\lambda_{ik}^a\bar{S}_{kj} + \lambda_{jk}^a\bar{S}_{ki}\right)\left(\lambda_{ik}^a S_{kj} + \lambda_{jk}^a S_{ki}\right) \\
&\quad \left. + g'^2 V^m V_m\bar{\rho}\rho + g'^2 V^m V_m\bar{\chi}\chi + 2gg'V_m^a V^m(\bar{\rho}\lambda^a\rho) - 2gg'V_m^a V^m(\bar{\chi}\lambda^a\chi)\right].
\end{aligned} \quad (A6)$$

### 4. Superpotential

$$\begin{aligned}
\mathcal{L}_{HMT} &= -\frac{\mu_\eta}{2}\tilde{\eta}_i\tilde{\eta}_i^* - \frac{\mu_\rho}{2}\tilde{\rho}_i\tilde{\rho}_i^* - \frac{\mu_\chi}{2}\tilde{\chi}_i\tilde{\chi}_i^* - \frac{\mu_S}{2}\tilde{S}_{ij}\tilde{S}_{ji}^* + hc. \\
\mathcal{L}_F &= \frac{1}{3}[3\lambda_1\epsilon F_L\tilde{L}\tilde{L} + \lambda_2\epsilon(2F_L\eta + F_\eta\tilde{L})\tilde{L} + \lambda_3(2F_L S + F_S\tilde{L})\tilde{L} \\
&\quad + f_1\epsilon(F_\rho\chi\eta + \rho F_\chi\eta + \rho\chi F_\eta) + f_2(2F_\eta\eta S + \eta\eta F_S) + f_3(F_\rho\chi S + \rho F_\chi S + \rho\chi F_S) + hc \\
\mathcal{L}_{H\tilde{H}\tilde{H}}^{W3} &= -\frac{1}{3}[f_1\epsilon(\tilde{\rho}\tilde{\chi}\eta + \rho\tilde{\chi}\tilde{\eta} + \tilde{\rho}\chi\tilde{\eta}) + f_2(\tilde{\eta}\tilde{\eta}S + \eta\tilde{\eta}\tilde{S} + \tilde{\eta}\eta\tilde{S}) + f_3(\tilde{\rho}\tilde{\chi}S + \rho\tilde{\chi}\tilde{S} + \tilde{\rho}\chi\tilde{S}) \\
&\quad + f_1'\epsilon(\tilde{\rho}'\tilde{\chi}'\eta' + \rho'\tilde{\chi}'\tilde{\eta}' + \tilde{\rho}'\chi'\tilde{\eta}') + f_2'(\tilde{\eta}'\tilde{\eta}'S' + \eta'\tilde{\eta}'\tilde{S}' + \tilde{\eta}'\eta'\tilde{S}') \\
&\quad + f_3'(\tilde{\rho}'\tilde{\chi}'S' + \rho'\tilde{\chi}'\tilde{S}' + \tilde{\rho}'\chi'\tilde{S}')] + hc.
\end{aligned} \quad (A7, A8)$$

### 5. Soft Term

$$\begin{aligned}
\mathcal{L}_{GMT}^{\text{gaugino}} &= -\frac{1}{2}\left[m_\lambda\sum_{a=1}^{8}(\lambda_A^a\lambda_A^a) + m'\lambda_B\lambda_B + hc\right], \\
\mathcal{L}_{\text{scalar}}^{\text{soft}} &= -m_\eta^2\bar{\eta}\eta - m_\rho^2\bar{\rho}\rho - m_\chi^2\bar{\chi}\chi - m_S^2\bar{S}S + (k_1\epsilon_{ijk}\rho_i\chi_j\eta_k + k_2\eta_i\eta_j\bar{S}_{ij} \\
&\quad + k_3\chi_i\rho_j\bar{S}_{ij} + hc), \\
\mathcal{L}_{SMT} &= -m_L^2\tilde{L}^\dagger\tilde{L} + \zeta_0\sum_{i=1}^{3}\sum_{j=1}^{3}\left(\tilde{L}_i\tilde{L}_j S_{ij} + \bar{\tilde{L}}_i\bar{\tilde{L}}_j S_{ij}^*\right).
\end{aligned} \quad (A9)$$

### APPENDIX B: NON DIAGONAL MASS MATRIX

In this Appendix we display all non diagonal mass matrix of the charginos and neutralinos.

### 1. Double Charged Chargino

Introducing the notation

$$\psi^{++} = \begin{pmatrix} -i\lambda_U^{++} & \tilde{\rho}^{++} & \tilde{\chi}'^{++} & \tilde{H}_1^{++} & \tilde{H}_2'^{++} \end{pmatrix}^t, \quad \psi^{--} = \begin{pmatrix} -i\lambda_U^{--} & \tilde{\rho}'^{--} & \tilde{\chi}^{--} & \tilde{H}_1'^{--} & \tilde{H}_2^{--} \end{pmatrix}^t,$$

and

$$\Psi^{\pm\pm} = \begin{pmatrix} \psi^{++} & \psi^{--} \end{pmatrix}^t, \tag{B1}$$

we can write Eq.(3.1) as follows

$$\mathcal{L}_{\text{mass}}^{\text{double}} = -\frac{1}{2}\left(\Psi^{\pm\pm}\right)^t Y^{\pm\pm}\Psi^{\pm\pm} + hc, \tag{B2}$$

where

$$Y^{\pm\pm} = \begin{pmatrix} 0 & T^t \\ T & 0 \end{pmatrix}, \tag{B3}$$

with

$$T = \begin{pmatrix} -m_\lambda & -gu & gw' & \frac{gz}{\sqrt{2}} & -\frac{gz'}{\sqrt{2}} \\ gu' & \frac{\mu_\rho}{2} & -\left(\frac{f_1'v'}{3} - \sqrt{2}f_3'z'\right) & 0 & f_3'w' \\ -gw & -\left(\frac{f_1 v}{3} - \sqrt{2}f_3 z\right) & \frac{\mu_\chi}{2} & f_3 u & 0 \\ -\frac{gz'}{\sqrt{2}} & 0 & f_3'u' & \frac{\mu_S}{2} & 0 \\ \frac{gz}{\sqrt{2}} & f_3 w & 0 & 0 & \frac{\mu_S}{2} \end{pmatrix}. \tag{B4}$$

The matrix $Y^{\pm\pm}$ in Eq.(3.2) satisfy the following relation

$$\det(Y^{\pm\pm} - \lambda I) = \det\left[\begin{pmatrix} -\lambda & T^t \\ T & -\lambda \end{pmatrix}\right] = \det(\lambda^2 - T^t \cdot T), \tag{B5}$$

so we only have to calculate $T^t \cdot T$ to obtain eigenvalues. Since $T^t \cdot T$ is a symmetric matrix, $\lambda^2$ must be real, and positive because $Y^{\pm\pm}$ is also symmetric.

### 2. Charged Chargino

Introducing the notation

$$\psi^+ = \begin{pmatrix} -i\lambda_W^+ & -i\lambda_V^+ & \tilde{\eta}_1'^+ & \tilde{\eta}_2^+ & \tilde{\rho}^+ & \tilde{\chi}'^+ & \tilde{h}_1'^+ & \tilde{h}_2^+ \end{pmatrix}^t, \quad \psi^- = \begin{pmatrix} -i\lambda_W^- & -i\lambda_V^- & \tilde{\eta}_1^- & \tilde{\eta}_2'^- & \tilde{\rho}'^- & \tilde{\chi}^- & \tilde{h}_1^- & \tilde{h}_2'^- \end{pmatrix}^t,$$

and

$$\Psi^\pm = \begin{pmatrix} \psi^+ & \psi^- \end{pmatrix}^t, \tag{B6}$$

Eq.(3.7) takes the form

$$\mathcal{L}_{\text{mass}}^{\text{unique}} = -\frac{1}{2}\left(\Psi^\pm\right)^t Y^\pm \Psi^\pm + hc, \tag{B7}$$

where

$$Y^\pm = \begin{pmatrix} 0 & X^t \\ X & 0 \end{pmatrix}, \tag{B8}$$

with

$$X = \begin{pmatrix} -m_\lambda & 0 & gv' & 0 & -gu & 0 & -\frac{gz'}{2} & 0 \\ 0 & -m_\lambda & 0 & -gv & 0 & gw' & 0 & -\frac{gz}{2} \\ -gv & 0 & \frac{\mu_\eta}{2} & 0 & -\frac{f_1 w}{3} & 0 & 0 & 0 \\ 0 & gv' & 0 & \frac{\mu_\eta}{2} & 0 & \frac{f'_1 u'}{3} & 0 & 0 \\ gu' & 0 & -\frac{f'_1 w'}{3} & 0 & \frac{\mu_\rho}{2} & 0 & 0 & 0 \\ 0 & -gw & 0 & \frac{f_1 u}{3} & 0 & \frac{\mu_\chi}{2} & 0 & 0 \\ -\frac{gz}{2} & 0 & 0 & \frac{f_3 w}{\sqrt{2}} & 0 & 0 & \frac{\mu_S}{2} & 0 \\ 0 & \frac{gz'}{2} & 0 & 0 & \frac{f'_3 u'}{\sqrt{2}} & 0 & 0 & \frac{\mu_S}{2} \end{pmatrix}. \tag{B9}$$

We can show that our matrix $X$ satisfy the same properties than $T$ ( Eq.(B5)).

### 3. Neutralinos

Introducing the notation

$$\Psi^0 = \begin{pmatrix} i\lambda_A^3 & i\lambda_A^8 & i\lambda_B & \tilde{\eta}^0 & \tilde{\eta}'^0 & \tilde{\rho}^0 & \tilde{\rho}'^0 & \tilde{\chi}^0 & \tilde{\chi}'^0 & \tilde{\sigma}_1^0 & \tilde{\sigma}_1'^0 & \tilde{\sigma}_2^0 & \tilde{\sigma}_2'^0 \end{pmatrix}^t,$$

Eq.(3.12) takes the following form:

$$\mathcal{L}_{\text{mass}}^{\text{neutralino}} = -\frac{1}{2} \left(\Psi^0\right)^t Y^0 \Psi^0 + hc, \tag{B10}$$

where

$$Y^0 = \begin{pmatrix} -m_\lambda & 0 & 0 & -\frac{gv}{\sqrt{2}} & \frac{gv'}{\sqrt{2}} & \frac{gu}{\sqrt{2}} & -\frac{gu'}{\sqrt{2}} & 0 & 0 & 0 & 0 & \frac{gz}{2\sqrt{2}} & -\frac{gz'}{2\sqrt{2}} \\ 0 & -m_\lambda & 0 & -\frac{gv}{\sqrt{6}} & \frac{gv'}{\sqrt{6}} & -\frac{gu}{\sqrt{6}} & \frac{gu'}{\sqrt{6}} & \frac{2}{\sqrt{6}}gw & -\frac{2}{\sqrt{6}}gw' & 0 & 0 & \frac{gz}{2\sqrt{6}} & -\frac{gz'}{2\sqrt{6}} \\ 0 & 0 & -m' & 0 & 0 & -\frac{g'u}{\sqrt{2}} & \frac{g'u'}{\sqrt{2}} & \frac{g'w}{\sqrt{2}} & -\frac{g'w'}{\sqrt{2}} & 0 & 0 & 0 & 0 \\ -\frac{gv}{\sqrt{2}} & -\frac{gv}{\sqrt{6}} & 0 & 0 & \frac{\mu_\eta}{2} & \frac{f_1 w}{3} & 0 & -\frac{f_1 u}{3} & 0 & 0 & 0 & 0 & 0 \\ \frac{gv'}{\sqrt{2}} & \frac{gv'}{\sqrt{6}} & 0 & \frac{\mu_\eta}{2} & 0 & 0 & \frac{f'_1 w'}{3} & 0 & -\frac{f'_1 u'}{3} & 0 & 0 & 0 & 0 \\ \frac{gu}{\sqrt{2}} & -\frac{gu}{\sqrt{6}} & -\frac{g'u}{\sqrt{2}} & \frac{f_1 w}{3} & 0 & 0 & \frac{\mu_\rho}{2} & \frac{f_1 v}{3}+\sqrt{2}f_3 z & 0 & 0 & 0 & \frac{f_3 w}{\sqrt{2}} & 0 \\ -\frac{gu'}{\sqrt{2}} & \frac{gu'}{\sqrt{6}} & \frac{g'u'}{\sqrt{2}} & 0 & \frac{f'_1 w'}{3} & \frac{\mu_\rho}{2} & 0 & 0 & \frac{f'_1 v'}{3}+\sqrt{2}f'_3 z' & 0 & 0 & 0 & -\frac{f'_3 w'}{\sqrt{2}} \\ 0 & \frac{2}{\sqrt{6}}gw & \frac{g'w}{\sqrt{2}} & -\frac{f_1 u}{3} & 0 & \frac{f_1 v}{3}+\sqrt{2}f_3 z & 0 & 0 & \frac{\mu_\chi}{2} & 0 & 0 & \frac{f_3 u}{\sqrt{2}} & 0 \\ 0 & -\frac{2}{\sqrt{6}}gw' & -\frac{g'w'}{\sqrt{2}} & 0 & -\frac{f'_1 u'}{3} & 0 & \frac{f'_1 v'}{3}+\sqrt{2}f'_3 z' & \frac{\mu_\chi}{2} & 0 & 0 & 0 & 0 & \frac{f'_3 u'}{\sqrt{2}} \\ 0 & 0 & 0 & 0 & 0 & 0 & 0 & 0 & 0 & 0 & \frac{\mu_S}{2} & 0 & 0 \\ 0 & 0 & 0 & 0 & 0 & 0 & 0 & 0 & 0 & \frac{\mu_S}{2} & 0 & 0 & 0 \\ \frac{gz'}{2\sqrt{2}} & \frac{gz}{2\sqrt{6}} & 0 & 0 & 0 & \frac{f_3 w}{\sqrt{2}} & 0 & \frac{f_3 u}{\sqrt{2}} & 0 & 0 & 0 & 0 & \frac{\mu_S}{2} \\ -\frac{gz'}{2\sqrt{2}} & -\frac{gz'}{2\sqrt{6}} & 0 & 0 & 0 & 0 & -\frac{f'_3 w'}{\sqrt{2}} & 0 & \frac{f'_3 u'}{\sqrt{2}} & 0 & 0 & \frac{\mu_S}{2} & 0 \end{pmatrix}.$$

(B11)

## APPENDIX C: CONNECTION BETWEEN THE TWO- AND FOUR- COMPONENT SPINORS

In this Appendix we show the procedure to write two-components spinors in terms of four-components spinors.

### 1. Weak Eigenstates

The weak interaction eigenstates are:

$$\tilde{U} = \begin{pmatrix} -i\lambda_U^{++} \\ i\bar{\lambda}_U^{--} \end{pmatrix}, \ \tilde{U}^c = \begin{pmatrix} -i\lambda_U^{--} \\ i\bar{\lambda}_U^{++} \end{pmatrix}, \ \tilde{T}_1^{++} = \begin{pmatrix} \tilde{\rho}^{++} \\ \bar{\tilde{\rho}}'^{--} \end{pmatrix}, \ \tilde{T}_1^{c++} = \begin{pmatrix} \tilde{\rho}'^{--} \\ \bar{\tilde{\rho}}^{++} \end{pmatrix},$$

$$\tilde{T}_2^{++} = \begin{pmatrix} \tilde{\chi}'^{++} \\ \bar{\tilde{\chi}}^{--} \end{pmatrix}, \ \tilde{T}_2^{c++} = \begin{pmatrix} \tilde{\chi}^{--} \\ \bar{\tilde{\chi}}'^{++} \end{pmatrix}, \ \tilde{S}_1^{++} = \begin{pmatrix} \tilde{H}_1^{++} \\ \bar{\tilde{H}}_1'^{--} \end{pmatrix}, \ \tilde{S}_1^{c++} = \begin{pmatrix} \tilde{H}_1'^{--} \\ \bar{\tilde{H}}_1^{++} \end{pmatrix},$$

$$\tilde{S}_2^{++} = \begin{pmatrix} \tilde{H}_2'^{++} \\ \bar{\tilde{H}}_2^{--} \end{pmatrix}, \ \tilde{S}_2^{c++} = \begin{pmatrix} \tilde{H}_2^{--} \\ \bar{\tilde{H}}_2'^{++} \end{pmatrix}, \ \tilde{W} = \begin{pmatrix} -i\lambda_W^+ \\ i\bar{\lambda}_W^- \end{pmatrix}, \ \tilde{W}^c = \begin{pmatrix} -i\lambda_W^- \\ i\bar{\lambda}_W^+ \end{pmatrix},$$

$$\tilde{V} = \begin{pmatrix} -i\lambda_V^+ \\ i\bar{\lambda}_V^- \end{pmatrix}, \ \tilde{V}^c = \begin{pmatrix} -i\lambda_V^- \\ i\bar{\lambda}_V^+ \end{pmatrix}, \ \tilde{T}_1^+ = \begin{pmatrix} \tilde{\eta}_1'^+ \\ \bar{\tilde{\eta}}_1^- \end{pmatrix}, \ \tilde{T}_1^{c+} = \begin{pmatrix} \tilde{\eta}_1^- \\ \bar{\tilde{\eta}}_1'^+ \end{pmatrix},$$

$$\tilde{T}_2^+ = \begin{pmatrix} \tilde{\eta}_2^+ \\ \bar{\tilde{\eta}}_2'^- \end{pmatrix}, \ \tilde{T}_2^{c+} = \begin{pmatrix} \tilde{\eta}_2'^- \\ \bar{\tilde{\eta}}_2^+ \end{pmatrix}, \ \tilde{T}_3^+ = \begin{pmatrix} \tilde{\rho}^+ \\ \bar{\tilde{\rho}}'^- \end{pmatrix}, \ \tilde{T}_3^{c+} = \begin{pmatrix} \tilde{\rho}'^- \\ \bar{\tilde{\rho}}^+ \end{pmatrix},$$

$$\tilde{T}_4^+ = \begin{pmatrix} \tilde{\chi}'^+ \\ \bar{\tilde{\chi}}^- \end{pmatrix}, \ \tilde{T}_4^{c+} = \begin{pmatrix} \tilde{\chi}^- \\ \bar{\tilde{\chi}}'^+ \end{pmatrix}, \ \tilde{S}_1^+ = \begin{pmatrix} \tilde{h}_1^- \\ \bar{\tilde{h}}_1'^+ \end{pmatrix}, \ \tilde{S}_1^{c+} = \begin{pmatrix} \tilde{h}_1'^+ \\ \bar{\tilde{h}}_1^- \end{pmatrix},$$

$$\tilde{S}_2^+ = \begin{pmatrix} \tilde{h}_2^+ \\ \bar{\tilde{h}}_2'^- \end{pmatrix}, \ \tilde{S}_2^{c+} = \begin{pmatrix} \tilde{h}_2'^- \\ \bar{\tilde{h}}_2^+ \end{pmatrix}, \ \tilde{W}_3 = \begin{pmatrix} -i\lambda_A^3 \\ i\bar{\lambda}_A^3 \end{pmatrix}, \ \tilde{W}_8 = \begin{pmatrix} -i\lambda_A^8 \\ i\bar{\lambda}_A^8 \end{pmatrix}, \ \tilde{B} = \begin{pmatrix} -i\lambda_B \\ i\bar{\lambda}_B \end{pmatrix},$$

$$\tilde{T}_1^0 = \begin{pmatrix} \tilde{\eta}^0 \\ \bar{\tilde{\eta}}^0 \end{pmatrix}, \ \tilde{T}_2^0 = \begin{pmatrix} \tilde{\eta}'^0 \\ \bar{\tilde{\eta}}'^0 \end{pmatrix}, \ \tilde{T}_3^0 = \begin{pmatrix} \tilde{\rho}^0 \\ \bar{\tilde{\rho}}^0 \end{pmatrix}, \ \tilde{T}_4^0 = \begin{pmatrix} \tilde{\rho}'^0 \\ \bar{\tilde{\rho}}'^0 \end{pmatrix}, \ \tilde{T}_5^0 = \begin{pmatrix} \tilde{\chi}^0 \\ \bar{\tilde{\chi}}^0 \end{pmatrix},$$

$$\tilde{T}_6^0 = \begin{pmatrix} \tilde{\chi}'^0 \\ \bar{\tilde{\chi}}'^0 \end{pmatrix}, \ \tilde{S}_1^0 = \begin{pmatrix} \tilde{\sigma}_1^0 \\ \bar{\tilde{\sigma}}_1^0 \end{pmatrix}, \ \tilde{S}_2^0 = \begin{pmatrix} \tilde{\sigma}_1'^0 \\ \bar{\tilde{\sigma}}_1'^0 \end{pmatrix},$$

$$\tilde{S}_3^0 = \begin{pmatrix} \tilde{\sigma}_2^0 \\ \bar{\tilde{\sigma}}_2^0 \end{pmatrix}, \ \tilde{S}_4^0 = \begin{pmatrix} \tilde{\sigma}_2'^0 \\ \bar{\tilde{\sigma}}_2'^0 \end{pmatrix}. \tag{C1}$$

With the states defined in Eqs.(C1) we get the following identities, that allow us to write the two-component spinors in terms of (four-component) weak eigenstates

$$\lambda_U^{--}\sigma^m\bar{\lambda}_A^3 = -\bar{\tilde{U}}L\gamma^m R\tilde{W}_3, \ \lambda_U^{++}\sigma^m\bar{\lambda}_A^3 = -\bar{\tilde{U}}^c L\gamma^m R\tilde{W}_3,$$

$$\lambda_A^8\sigma^m\bar{\lambda}_U^{++} = -\bar{\tilde{W}}_8 L\gamma^m R\tilde{U}^c, \ \lambda_U^{++}\sigma^m\bar{\lambda}_A^8 = -\bar{\tilde{U}}^c L\gamma^m R\tilde{W}_8,$$

$$\tilde{\chi}'^{++}\sigma^m\bar{\tilde{\chi}}'^0 = -\bar{\tilde{T}}_3^{c++} L\gamma^m R\tilde{T}_6^0, \ \tilde{\rho}^{++}\sigma^m\bar{\tilde{\rho}}^0 = -\bar{\tilde{T}}_1^{c++} L\gamma^m R\tilde{T}_3^0,$$

$$\tilde{\chi}^0\sigma^m\bar{\tilde{\chi}}^{--} = -\bar{\tilde{T}}_5^0 L\gamma^m R\tilde{T}_2^{++}, \ \tilde{\rho}'^0\sigma^m\bar{\tilde{\rho}}'^{--} = -\bar{\tilde{T}}_4^0 L\gamma^m R\tilde{T}_1^{++},$$

$$\tilde{\sigma}_2'^0\sigma^m\bar{\tilde{H}}_1'^{--} = -\bar{\tilde{S}}_4^0 L\gamma^m R\tilde{S}_1^{++}, \ \tilde{H}_1^{++}\sigma^m\bar{\tilde{\sigma}}_2^0 = -\bar{\tilde{S}}_1^{c++} L\gamma^m R\tilde{S}_3^0,$$

$$\tilde{H}_2'^{++}\sigma^m\bar{\tilde{\sigma}}_2'^0 = -\bar{\tilde{S}}_2^{c++} L\gamma^m R\tilde{S}_4^0, \ \tilde{\sigma}_2^0\sigma^m\bar{\tilde{H}}_2^{--} = -\bar{\tilde{S}}_3^0 L\gamma^m R\tilde{S}_2^{++},$$

$$\lambda_V^-\sigma^m\bar{\lambda}_W^+ = -\bar{\tilde{V}}L\gamma^m R\tilde{W}^c, \ \lambda_W^+\sigma^m\bar{\lambda}_V^- = -\bar{\tilde{W}}^c L\gamma^m R\tilde{V},$$

$$\tilde{\eta}_2^+\sigma^m\bar{\tilde{\eta}}_1^- = -\bar{\tilde{T}}_2^{c+} L\gamma^m R\tilde{T}_1^+, \ \tilde{h}_1'^+\sigma^m\bar{\tilde{h}}_2'^- = -\bar{\tilde{S}}_1^{c+} L\gamma^m R\tilde{S}_2^+,$$

$$\tilde{\eta}_1'^+\sigma^m\bar{\tilde{\eta}}_2'^- = -\bar{\tilde{T}}_1^{c+} L\gamma^m R\tilde{T}_2^+, \ \tilde{h}_2^+\sigma^m\bar{\tilde{h}}_1^- = -\bar{\tilde{S}}_2^{c+} L\gamma^m R\tilde{S}_1^+. \tag{C2}$$

### 2. Mass Eigenstates

From Eq.(3.3, 3.9, 3.15), we can write the inverse transformation as:

$$\Psi_k^{++} = A_{ik}^*\tilde{\chi}_i^{++}, \ \Psi_k^{--} = B_{ik}^*\tilde{\chi}_i^{--}, \ \Psi_k^+ = D_{ik}^*\tilde{\chi}_i^+, \ \Psi_k^- = E_{ik}^*\tilde{\chi}_i^-,$$
$$\Psi_k^0 = N_{ik}^*\tilde{\chi}_i^0. \tag{C3}$$

Equations(C2) above do not involve physical particles. From Eqs.(3.6) and (3.16) we can show the following

$$\tilde{\chi}_i^{++} = L\Psi(\tilde{\chi}_i^{++}), \ \bar{\tilde{\chi}}_i^{++} = R\Psi^c(\tilde{\chi}_i^{--}), \ \bar{\tilde{\chi}}_i^{--} = R\Psi(\tilde{\chi}_i^{++}), \ \tilde{\chi}_i^{--} = L\Psi^c(\tilde{\chi}_i^{--}),$$
$$\tilde{\chi}_i^0 = L\Psi(\tilde{\chi}_i^0), \ \bar{\tilde{\chi}}_i^0 = R\Psi(\tilde{\chi}_i^0),$$
$$\tilde{\chi}_i^+ = L\Psi(\tilde{\chi}_i^+), \ \bar{\tilde{\chi}}_i^+ = R\Psi^c(\tilde{\chi}_i^-), \ \bar{\tilde{\chi}}_i^- = R\Psi(\tilde{\chi}_i^+), \ \tilde{\chi}_i^- = L\Psi^c(\tilde{\chi}_i^-). \tag{C4}$$

### APPENDIX D: DIFFERENTIAL CROSS SECTION

In this Appendix we calculate the differential cross section to the processes we have studied in section VI.

**1.** $e^-e^- \to \tilde{\chi}^-\tilde{\chi}^-$ ( **Charginos productions**)

$$\frac{d\sigma}{d\Omega}(e^-e^- \to \tilde{\chi}^-\tilde{\chi}^-) = \frac{1}{128\pi^2 s}\sqrt{\frac{s}{\frac{s}{4} - m_{\tilde{\chi}^+}^2}}|\mathcal{M}_T|^2, \tag{D1}$$

where

$$|\mathcal{M}_T|^2 = |D_{i7}|^4 \left(\frac{1}{(t - m_{\tilde{\nu}}^2)^2} + \frac{1}{(u - m_{\tilde{\nu}}^2)^2}\right) \left\{2m_{\tilde{\chi}^+}^4 + 2(E - 2m_{\tilde{\chi}^+}^2)m_{\tilde{\chi}^+}^2\right.$$
$$\left. + 2\left[(m_{\tilde{\chi}^+}^2 - \frac{s}{2})^2 + s\left(\frac{s}{4} - m_{\tilde{\chi}^+}^2\right)\cos\theta\right]\right\}$$
$$- \frac{2g^2(s - M_U^2)O_2|D_{i7}|^2}{(s - M_U^2)^2 + (\Gamma_U M_U)^2}\left[m_{\tilde{\chi}^+}^4\left(\frac{1}{(t - m_{\tilde{\nu}}^2)} + \frac{1}{(u - m_{\tilde{\nu}}^2)}\right)\right.$$
$$\left. + (2m_{\tilde{\chi}^+}^2 + s)\left(\frac{u}{(t - m_{\tilde{\nu}}^2)} + \frac{t}{(u - m_{\tilde{\nu}}^2)}\right)\right],$$
$$+ \frac{g^4|O_2|^2}{(s - M_U^2)^2 + (\Gamma_U M_U)^2}(m_{\tilde{\chi}^+}^4 - 8sm_{\tilde{\chi}^+}^2 + 4s^2) + \frac{4s|D_{i7}|^2}{(t - m_{\tilde{\nu}}^2)(u - m_{\tilde{\nu}}^2)}.$$

**2.** $e^-e^- \to \tilde{\chi}^{--}\tilde{\chi}^0$ ( **Double charged Neutralinos production**)

$$\frac{d\sigma}{d\Omega}(e^-e^- \to \tilde{\chi}^{--}\tilde{\chi}^0) = \frac{1}{64\pi^2 s}\sqrt{\frac{s}{E_{\chi^{++}}^2 - m_{\chi^{++}}^2}}|\mathcal{M}_T|^2, \tag{D2}$$

where

$$|\mathcal{M}_T|^2 = \left(\frac{X_1^2 \cos^2\theta_f}{(u - m_{\tilde{l}_2}^2)^2} + \frac{X_2^2 \cos^2\theta_f}{(t - m_{\tilde{l}_2}^2)^2} + \frac{X_3^2 \sin^2\theta_f}{(t - m_{\tilde{l}_1}^2)^2} + \frac{X_4^2 \sin^2\theta_f}{(u - m_{\tilde{l}_1}^2)^2}\right.$$
$$\left. + \frac{X_1 X_4 \sin\theta_f \cos\theta_f}{(u - m_{\tilde{l}_2}^2)(u - m_{\tilde{l}_1}^2)} + \frac{X_2 X_3 \sin\theta_f \cos\theta_f}{(t - m_{\tilde{l}_2}^2)(t - m_{\tilde{l}_1}^2)}\right)$$
$$\cdot [2(m_{\chi^{++}}^4 + m_{\chi^0}^4) + s(m_{\chi^{++}}^2 + m_{\chi^0}^2 - s) + 2t(t - 2m_{\chi^{++}}^2) + 2u(u - 2m_{\chi^0}^2)]$$
$$+ 2m_{\chi^{++}}^4 m_{\chi^0}^4 O^1 \cos\theta_f \left(\frac{X_1}{(u - m_{\tilde{l}_2}^2)^2[(s - M_U^2)^2 + (\Gamma_U M_U)^2]}\right.$$
$$\left. + \frac{X_2}{(t - m_{\tilde{l}_2}^2)^2[(s - M_U^2)^2 + (\Gamma_U M_U)^2]}\right)$$
$$+ 2m_{\chi^{++}}^4 m_{\chi^0}^4 O^1 \sin\theta_f \left(\frac{X_3}{(t - m_{\tilde{l}_1}^2)^2[(s - M_U^2)^2 + (\Gamma_U M_U)^2]}\right.$$
$$\left. + \frac{X_4}{(u - m_{\tilde{l}_1}^2)^2[(s - M_U^2)^2 + (\Gamma_U M_U)^2]}\right)$$
$$+ \frac{O^1}{(s - M_U^2)^2 + (\Gamma_U M_U)^2}[2(m_{\chi^{++}}^4 + m_{\chi^0}^2) + 4(2(m_{\chi^{++}}^2 + m_{\chi^0}^2))s + s^2],$$

where

$$X_1 = \frac{2}{\sqrt{2}} A_{i5}\lambda_3 N_{i8}, \quad X_2 = \frac{2}{\sqrt{2}} A_{i1} g N_{i8},$$
$$X_3 = A_{i1} g \left(\frac{N_{i1}}{\sqrt{2}} + \frac{N_{i2}}{\sqrt{6}}\right), \quad X_4 = A_{i5}\lambda_3 \left(\frac{N_{i1}}{\sqrt{2}} + \frac{N_{i2}}{\sqrt{6}}\right). \tag{D3}$$

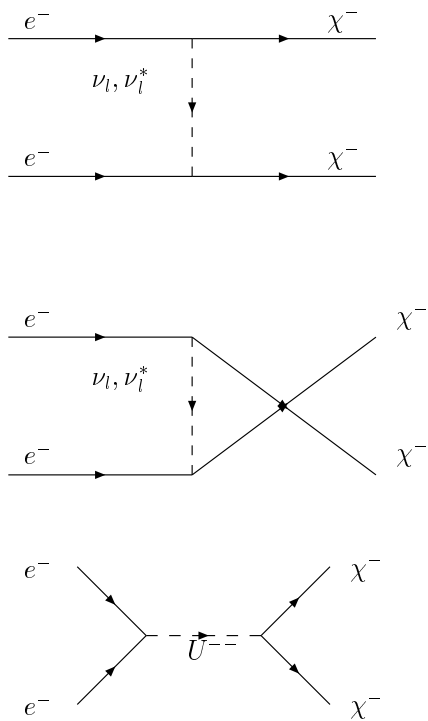

FIG. 1. Feynman diagrams for the process $e^- e^- \to \tilde{\chi}^- \tilde{\chi}^-$.

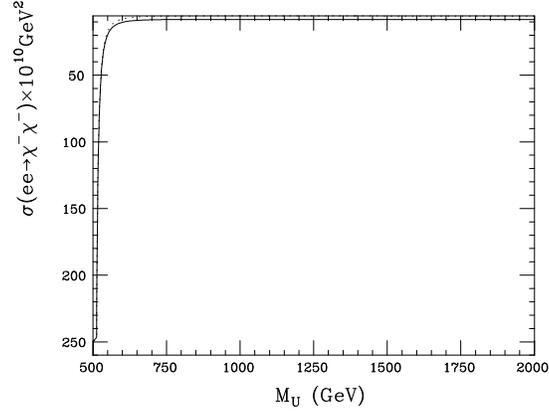

FIG. 2. Total Cross Section $e^-e^- \to \tilde{\chi}^-\tilde{\chi}^-$ at $\sqrt{s} = 0.5\text{TeV}$ and $O_2 = 1$, $A_{i1} = 10^{-1}$.

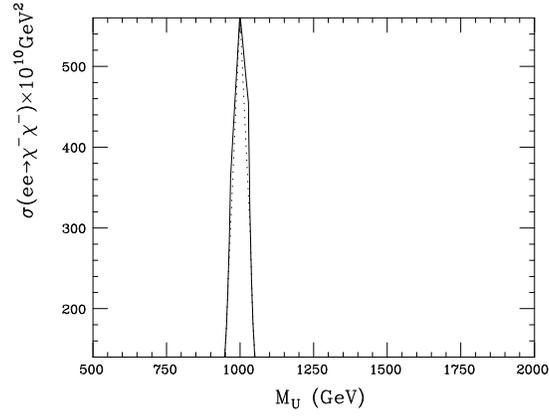

FIG. 3. Total Cross Section $e^-e^- \to \tilde{\chi}^-\tilde{\chi}^-$ at $\sqrt{s} = 1.0\text{TeV}$ and $D_{i7} = 10^{-1}$, $O_2 = 10^{-1}$.

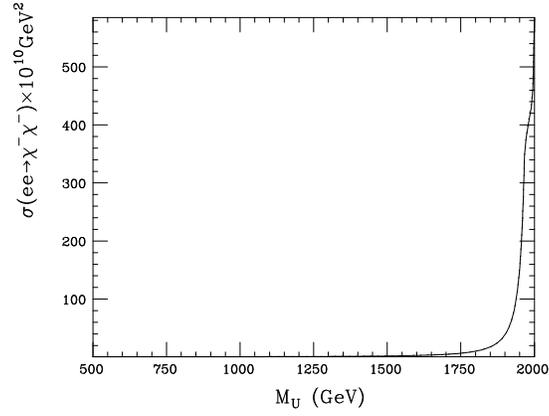

FIG. 4. Total Cross Section $e^-e^- \to \tilde{\chi}^-\tilde{\chi}^-$ at $\sqrt{s} = 2.0\text{TeV}$ and $D_{i7} = 1$, $O_2 = 10^{-1}$.

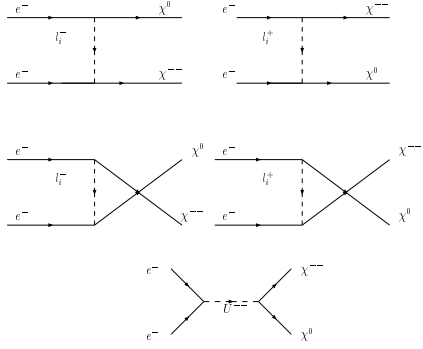

FIG. 5. Feynman diagrams for the process $e^-e^- \to \tilde{\chi}^{--}\tilde{\chi}^0$, and $i = 1, 2$.

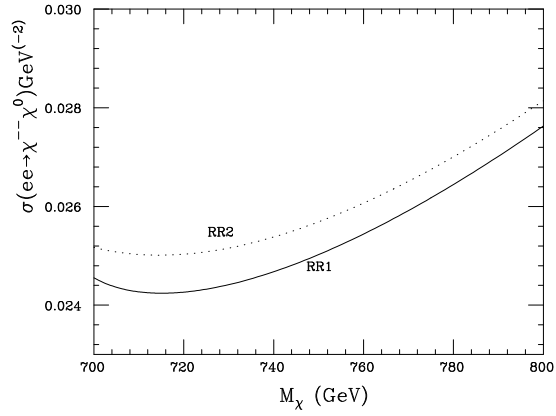

FIG. 6. Total Cross Section $e^-e^- \to \tilde{\chi}^{--}\tilde{\chi}^0$ at $\sqrt{s} = 0.5\text{TeV}$ and $X_1 \cos\theta_f = X_2 \cos\theta_f = X_3 \sin\theta_f = X_4 \sin\theta_f = 10^{-1}$, $O^1 = 10^{-2}$.

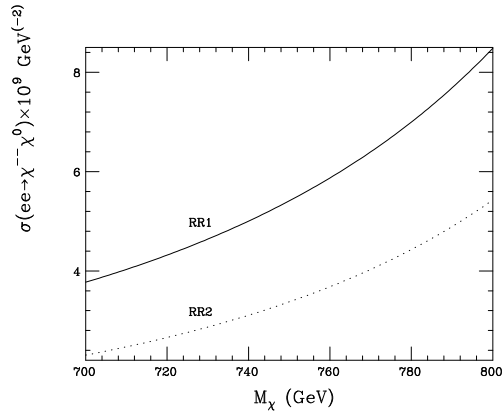

FIG. 7. Total Cross Section $e^-e^- \to \tilde{\chi}^{--}\tilde{\chi}^0$ at $\sqrt{s} = 1.0\text{TeV}$ and $X_1 \cos\theta_f = X_2 \cos\theta_f = X_3 \sin\theta_f = X_4 \sin\theta_f = 10^{-1}$, $O^1 = 10^{-1}$.

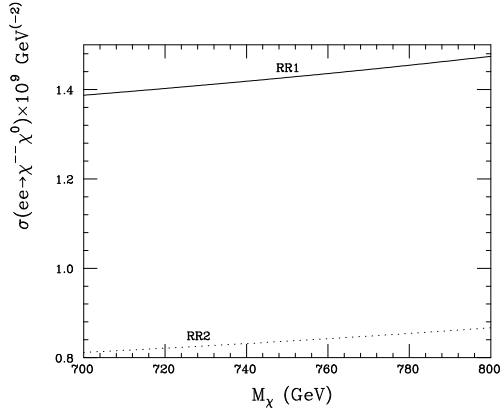

FIG. 8. Total Cross Section $e^-e^- \to \tilde{\chi}^{--}\tilde{\chi}^0$ at $\sqrt{s} = 2.0$TeV and
$X_1 \cos\theta_f = X_2 \cos\theta_f = X_3 \sin\theta_f = X_4 \sin\theta_f = 10^{-2}$, $O^1 = 10^{-2}$.

| Vertices | Feynman rules |
|---|---|
| $l^-l^-U^{--}$ | $-\frac{ig}{\sqrt{2}}C\gamma^m L$ |
| $\tilde{\chi}_j^{--}\tilde{\chi}_i^0 U^{--}$ | $\frac{ig}{2}O_{ij}^1 C\gamma^m R$ |
| $\tilde{\chi}_i^-\tilde{\chi}_j^- U^{--}$ | $\frac{ig}{2}O_{ij}^2 C\gamma^m R$ |
| $\tilde{l}_1^- l^- \tilde{\chi}_i^{--}$ | $-2i\lambda_3 A_{i5} \sin\theta_f R$ |
| $\tilde{l}_2^- l^- \tilde{\chi}_i^{--}$ | $-2i\lambda_3 A_{i5} \cos\theta_f R$ |
| $\tilde{l}_1^- l^- \tilde{\chi}_i^0$ | $i\left[g\left(\frac{N_{i1}}{\sqrt{2}} + \frac{N_{i2}}{\sqrt{6}}\right)\cos\theta_f R - \lambda_3 \frac{2}{\sqrt{2}}\sin\theta_f N_{i8} R\right]$ |
| $\tilde{l}_2^- l^- \tilde{\chi}_i^0$ | $i\left[g\left(\frac{N_{i1}}{\sqrt{2}} + \frac{N_{i2}}{\sqrt{6}}\right)\sin\theta_f R + \lambda_3 \frac{2}{\sqrt{2}}\cos\theta_f N_{i8} R\right]$ |
| $\tilde{l}_1^+ l^- \tilde{\chi}_i^0$ | $i\left[g\left(\frac{N_{i1}^*}{\sqrt{2}} + \frac{N_{i2}^*}{\sqrt{6}}\right)\cos\theta_f L - \lambda_3 \frac{2}{\sqrt{2}}\sin\theta_f N_{i8}^* L\right]$ |
| $\tilde{l}_2^+ l^- \tilde{\chi}_i^0$ | $i\left[g\left(\frac{N_{i1}^*}{\sqrt{2}} + \frac{N_{i2}^*}{\sqrt{6}}\right)\sin\theta_f L + \lambda_3 \frac{2}{\sqrt{2}}\cos\theta_f N_{i8}^* L\right]$ |
| $\tilde{l}_1^+ l^- \tilde{\chi}_i^{--}$ | $-igA_{i1}\sin\theta_f RC$ |
| $\tilde{l}_2^+ l^- \tilde{\chi}_i^{--}$ | $-igA_{i1}\cos\theta_f RC$ |
| $\tilde{\nu}_l l^- \tilde{\chi}_i^-$ | $-i\lambda_3 \frac{2}{\sqrt{2}} D_{i7}^* L$ |
| $\tilde{\nu}_l^* l^- \tilde{\chi}_i^-$ | $-i\lambda_3 \frac{2}{\sqrt{2}} D_{i7} R$ |

TABLE I. Feynman rules derived from Eqs. (4.5,4.6,4.7,4.8).

| $\tilde{m}$ [GeV] | RR1 : $\tan\beta = 3$ | RR2 : $\tan\beta = 30$ |
|---|---|---|
| $\tilde{\chi}_1^\pm$ | 128 | 132 |
| $\tilde{\chi}_1^0$ | 70 | 72 |
| $\tilde{e}_L^-$ | 176 | 217 |
| $\tilde{e}_R^-$ | 132 | 183 |
| $\tilde{\nu}$ | 166 | 206 |

TABLE II. Mass values of the ligthest chargino and neutralino, sleptons and of the sneutrino at electro weak scale corresponding to the mSUGRA solution.

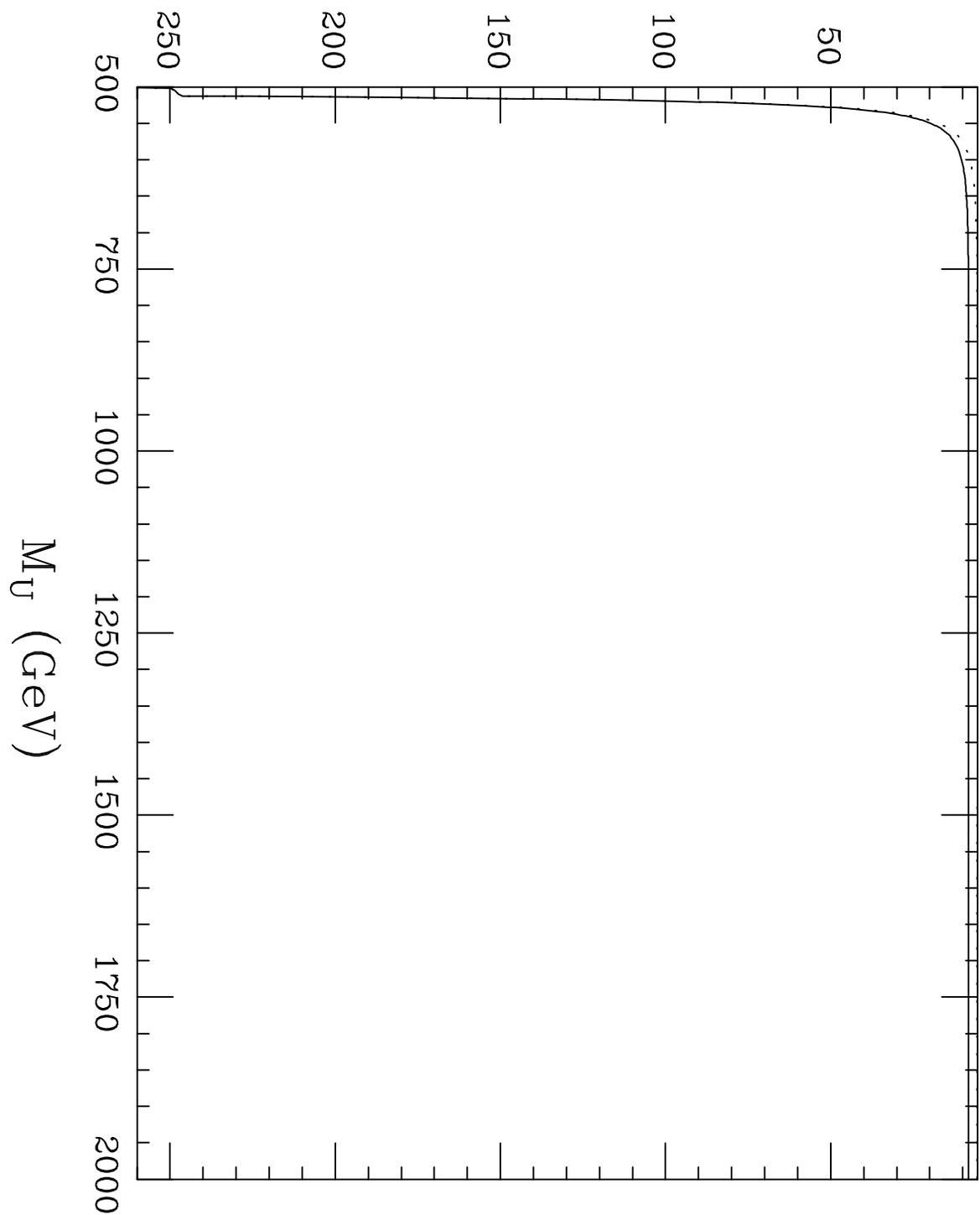

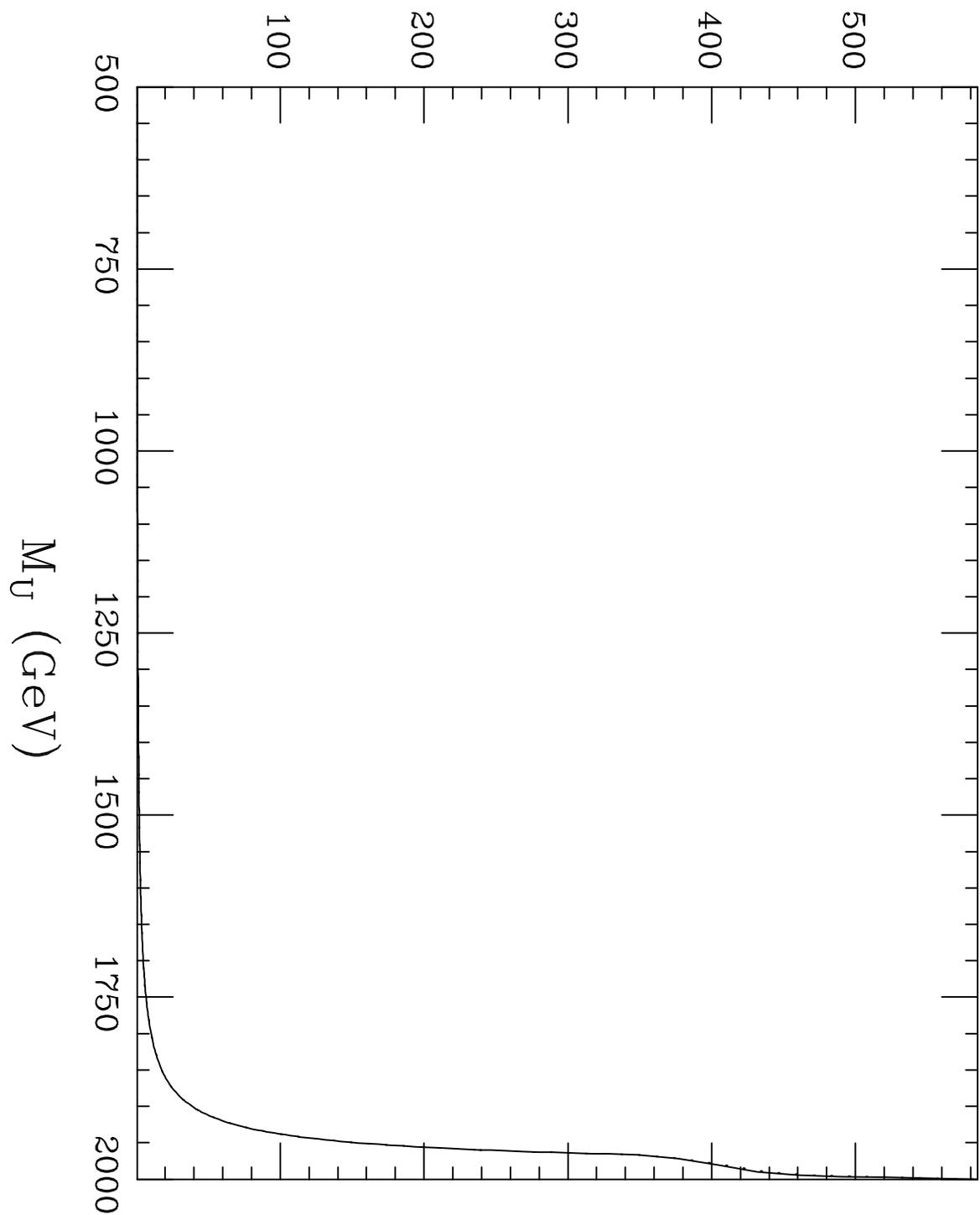

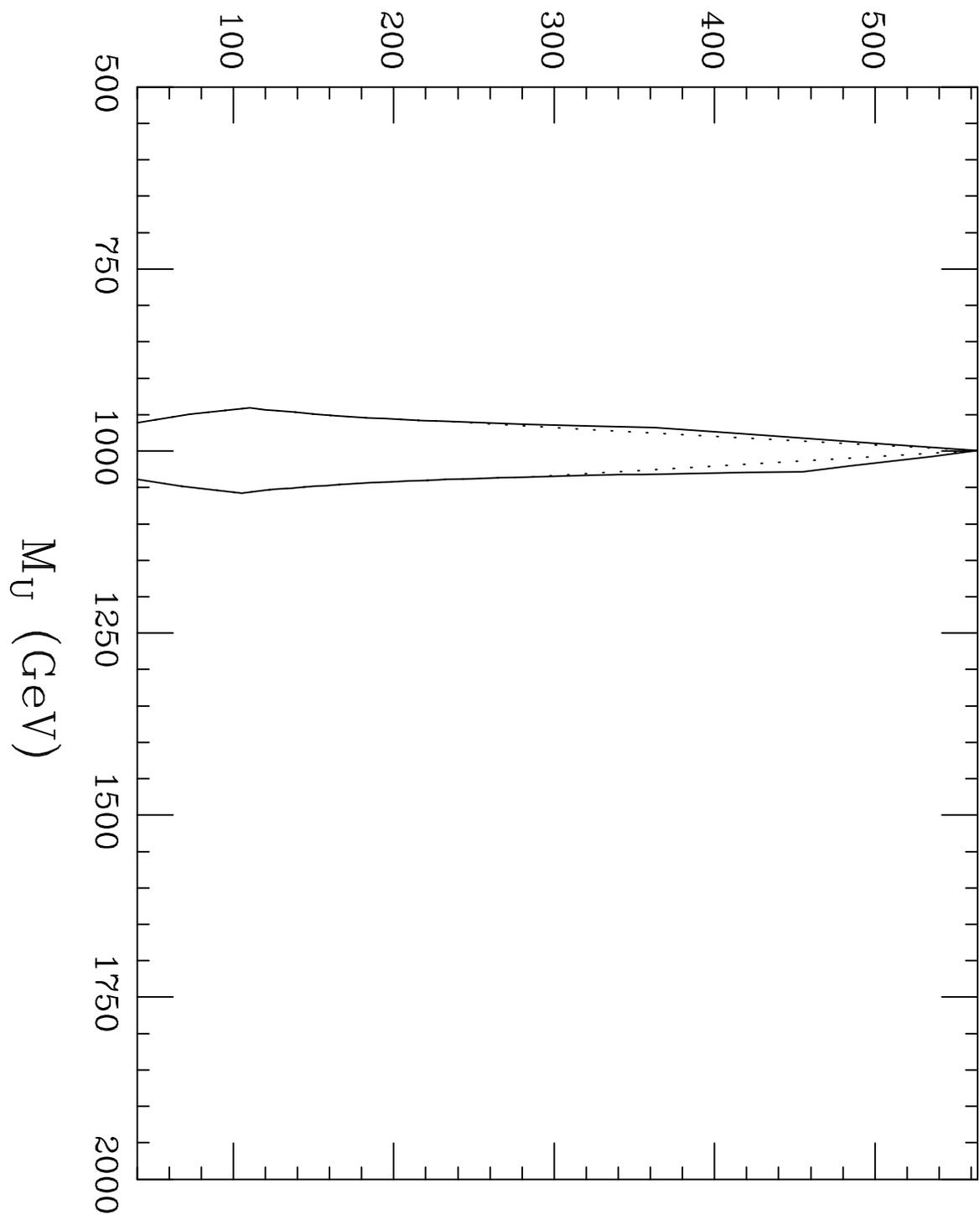